\documentclass[12pt]{article}
\pdfoutput=1
\usepackage{graphics}
\usepackage{amsmath}
\usepackage{makeidx}

\usepackage{a4,color,graphics}
\usepackage{amsfonts}
\usepackage{amssymb}
\usepackage{graphicx}%
\usepackage[all]{xy} 

\makeatletter \@addtoreset{equation}{section}

\def\be{\begin{equation}}
\def\ee{\end{equation}}
\def\bea{\begin{eqnarray}}
\def\eea{\end{eqnarray}}

\newcommand{\nc}{\newcommand}
\nc{\al}{\alpha} \nc{\bib}{\bibitem} \nc{\la}{\lambda}
\nc{\C}{\mbox{\hspace{1.24mm}\rule{0.2mm}{2.5mm}\hspace{-2.7mm}
C}} \nc{\R}{\mbox{\hspace{.04mm}\rule{0.2mm}{2.8mm}\hspace{-1.5mm}
R}}

\begin{document} 
\title{%
\textbf{
Bilocal fields and gravity 
}} 
\author{Pablo Diaz\thanks{pablo.diazbenito@uleth.ca}, 
~Saurya Das\thanks{saurya.das@uleth.ca}, ~Mark Walton\thanks{walton@uleth.ca} 
\\
{\small \emph{Theoretical Physics Group and Quantum Alberta, }}\\
{\small \emph{Department of Physics and Astronomy, University of Lethbridge,}}\\
{\small \emph{4401 University Drive, Lethbridge, Alberta, T1K 3M4, Canada}}
}

\maketitle
\begin{abstract}
We study a classical bilocal field theory perturbatively up to second order. 
The chosen theory is the simplest which 
incorporates action-at-a-distance, while keeping non-local effects short-ranged. 
We show that the new degrees of freedom introduced by bilocality can be 
interpreted as gravitational degrees of freedom in the following sense:
solutions of the bilocal system at linear and second order 
contain as a subset, gravitational perturbations (spacetime fluctuations) also
to that order. In other words, gravity can be thought to originate in a bilocal
field theory. We examine potential implications. 
\\

\textbf{Keywords}: Gravity, bilocal field, non-locality, perturbative gravity, gravitational waves.
\end{abstract}
\thispagestyle{empty} 
\newpage
\pagestyle{plain} 

\tableofcontents
\newpage

\section{Introduction}

It has been convincingly argued by many authors that there is a theoretical 
(an practical) impossibility of resolving two close by spacetime points beyond a minimum length scale, often chosen as the Planck scale \cite{gross,acv,maggiore,DFE}. 
This in turn suggests a fundamental non-locality of spacetime \cite{Ah}.
This idea is supported by the fact that many fundamental theories of quantum 
gravity are also non-local, most notable being string theory, but also 
spacetime foam, fuzzy spaces or non-commutative spacetime \cite{BBS,R,M,LY}. 
It has also been conjectured for a long time that gravity is a
low energy emergent phenomenon (e.g. \cite{emergent}). 
Therefore, it is natural to ask whether gravity can arise 
as an effective theory from an underlying non-local field theory, even at the classical level. 
This possibility was first considered in \cite{DDW} and is further explored in this article.

Since non-locality violates causality, one of the basic tenets of relativity, 
and verified to a high degree of accuracy to distance scales of 
about $10^{-19}~m\approx (1~TeV)^{-1}$,
one requires any reasonable non-local theory to show non-local effects at shorter distances,
and also controlled by a tunable parameter which in a certain precise limit renders
the theory local. 
Subject to this requirement, several possibilities exist. 
For example, 
one can consider theories defined on fuzzy geometries instead of manifolds, with 
a parameter $N$ which makes the theory local in the large $N\rightarrow\infty$ limit\footnote{In the fuzzy sphere case, $N$ parametrizes the dimension of the irreps of $SU(2)$ whose matrices play the role of spacetime (non-commutative) variables.} \cite{M}.
A second way of introducing non-locality is to have fields depending on 
more than one spacetime point. These so-called multilocal theories can be
bilocal, trilocal, tetralocal and so on. We explore
this option in this paper, in particular bilocal field theories.
Our bilocality is genuine in the sense of action-at-a-distance, and as mentioned
earlier, short-ranged. Be aware that our analysis is not restricted to bilocal theories but will also apply to 
non-local theories which lead to an effective 
bilocality when suitable degrees of freedom  are integrated out. 
In our model the scale $\alpha$, with units of mass, is the `non-local parameter',
with $\alpha \rightarrow \infty$ recovering locality%
\footnote{When comparing to gravity, we assume $\alpha\sim G^{-1/2}$, where $G$ is the Newton constant.}. 

In an earlier paper \cite{DDW}, we studied a bilocal model of this type 
(without self-interactions) and found that their massless solutions matched gravity waves. 
In this article we solve a bilocal model with a quadratic vertex and we compare its massless solutions with those of gravity perturbatively. We show that at linear order it reproduces gravity waves, see equation (\ref{goodsols}), and the second order, the bilocal contribution encodes second order contributions of gravity in the Lorentz gauge, cf. the relations (\ref{secondorderrelations}). We conjecture that what appears to be this 
straightforward yet highly non-trivial correspondence to hold to 
higher orders as well. We work in four dimensional spacetime, although there seems to be no obstruction in generalizing our analysis to an arbitrary dimension. 

Several points may be noted here. 
First, one is describing a spin-2 local field theory by means of a 
a specific spin-0 bilocal theory, which although is minimal in some sense, is
not unique. 
Second, the equations and interaction terms do not at all look similar, and as mentioned 
above, gravity solutions are reproduced in the specific Lorentz gauge; other bilocal
theories will presumably yield gravity in other gauges. 
Third, bilocal field theories of
this type were first studied by Yukawa in a series of papers, with the aim of describing
mesonic excitations, with the two spacetime points the field depends on being 
interpreted as the location of the quarks \cite{Yu}. Similarly,
the use of bilocal and trilocal fields to explain confinement in hadrons became popular in the next couple of decades \cite{T,bilocal,Fe}, until the success of quantum chromodynamics made them go away.  

In the context of the  AdS/CFT correspondence (in the higher-spin version), bilocal fields have also been reconsidered by Jevicki {\it et al.} during the last decade \cite{jevicki1,jevicki,dMJJR}. 
In those papers, the use of the collective field, which is bilocal, in the gauge side of the correspondence was proven to be crucial for describing the physical degrees of freedom of gravity and high spin fields in the gravity side. In fact, some of the ideas contained in those papers have inspired us, especially the prominent role of the bilocal field in the description of gravity. All along this paper we avoid (on purpose) a detailed connection of our approach to holography, since we would like to stress
that the dynamics of the bilocal field is in itself rich enough to encode the dynamics of a gravitational and higher spin fields. Nevertheless, the connection of our approach to holography and to the earlier works above-mentioned is clarified in section \ref{CWH}. 

The paper is organized as follows. To successfully compare 
bilocal theory with gravity, we start by reviewing some basics of perturbation theory in section \ref{permeth}. In section \ref{solgrav} we review the solutions for linear gravity and the second order contributions in the Lorentz gauge. Section \ref{BM} contains the main results of the paper. In this section we introduce the bilocal field and the bilocal model equations. We solve the model for the linear and the second order in subsections \ref{LO} and \ref{SO} respectively. In section \ref{GfB} we show how to ``extract'' the perturbative gravity solutions from the bilocal ones. 
Finally, we summarize our results and comment on future directions with this idea.

\section{Revision of the perturbative method}
\label{permeth}

Since we will compare 
bilocal theory and gravity perturbatively in this paper,
we briefly review the perturbative method in this section. 

\subsection{Perturbative models}

As it is customary in perturbation theory, we assume that there exist a one-parameter family of solutions $\Phi(\kappa_B)$, where $\kappa_B$ is a parameter of the theory. We expand the solution in powers of the (small) parameter $\kappa_B$ as
\begin{equation}\label{phiexp}
 \Phi(\kappa_B)=\Phi^{(1)}+\kappa_B\Phi^{(2)}+\kappa_B^2\Phi^{(3)}\dots~.
\end{equation}
We write the bilocal model as 
\begin{equation}\label{Ophi}
\mathcal{O}\Phi(\kappa_B)=V\big(\Phi(\kappa_B),\kappa_B\big),
\end{equation}
%
where $\mathcal{O}$ is a linear differential operator and $V(\Phi)$ stands for self-interactions, also expanded in powers of $k_B$. 
%
Then, equating the same powers of $k_B$ on both sides of Eq.(\ref{Ophi}), one gets an 
infinite set of differential equations, one for each $\Phi^{(i)}$, all linear, and each of them sourced by lower order solutions. These equations can be solved in ascending order, with first equation being
\begin{equation*}
\mathcal{O}\Phi^{(1)}=0.
\end{equation*}
The method we have just sketched is quite general. In section  \ref{BM} we will give a precise form of the differential operator $\mathcal{O}$ and the terms in the expansion of $V(\Phi)$ for the bilocal model we use. 

\subsection{Perturbative gravity}

In this case,
we start by assuming the existence of a one-parameter family of solutions $g(\kappa)_{\mu\nu}$ to Einstein's equations
\begin{equation*}\label{Einstein}
 R_{\mu\nu}-\frac{1}{2}g_{\mu\nu}R+\Lambda g_{\mu\nu}=8\pi G T_{\mu\nu}.
\end{equation*}
The perturbation equations are generated as the Einstein's equations are expanded in powers of $\kappa$ and the terms with equal powers are equated. 
In this paper, for simplicity, we will consider pure gravity with no cosmological constant, i.e. 
\begin{equation*}\label{Einsteinflat}
G_{\mu\nu}\equiv  R_{\mu\nu}-\frac{1}{2}g_{\mu\nu}R =0~.
\end{equation*}
Now, both the solutions and the `operator' $G_{\mu\nu}$ are 
expanded in powers of $\kappa$. Any solution of the $\kappa$-family
is expanded around the flat metric $\eta_{\mu\nu}$ as
\footnote{The parameter dimensional parameter 
$\kappa$ is chosen such that 
$\kappa=\sqrt{16\pi G}$, where $G$ is the Newton constant.} 
\begin{equation}
\label{solexp}
g_{\mu\nu}(\kappa)=\eta_{\mu\nu}+\kappa h^{(1)}_{\mu\nu}+\kappa^2  h^{(2)}_{\mu\nu}+\cdots ,
\end{equation}
where $h^{(i)}_{\mu\nu}$ are order $i$ contributions to the solution $g_{\mu\nu}(\kappa)$ and are found as solutions of some linear differential equations when solved in ascending order. To find out the tower of equations that $h^{(i)}_{\mu\nu}$  solves, we expand the Einstein's tensor. This is done by writing
\begin{equation}\label{opexp}
G_{\mu\nu}[\eta_{ab}+\kappa h_{ab}]=G^{(0)}_{\mu\nu}[\eta_{ab}]+\kappa G^{(1)}_{\mu\nu}[h_{ab}]+\kappa^2 G^{(2)}_{\mu\nu}[h_{ab}]+\cdots.
\end{equation}
A general order operator is then computed as 
\begin{equation*}
 G^{(n)}_{\mu\nu}[h_{ab}]=\frac{1}{n!}\frac{d^n}{d\kappa^n}G_{\mu\nu}[\eta_{ab}+\kappa h_{ab}]\bigg|_{\kappa=0}.
\end{equation*}
Performing expansions (\ref{solexp}) and (\ref{opexp}) and equating equal powers of $\kappa$ we get the tower of equations
\begin{eqnarray}
0&=& G^{(0)}_{\mu\nu}[\eta_{ab}] \nonumber 
\\
0&=& G^{(1)}_{\mu\nu}[h^{(1)}_{ab}], \label{linear} \\
0&=& G^{(1)}_{\mu\nu}[h^{(2)}_{ab}]+ G^{(2)}_{\mu\nu}[h^{(1)}_{ab}],\label{secondorder}\\
&\vdots& \nonumber \label{tower}
\end{eqnarray}
To find $h^{(2)}_{ab}$ one must solve (\ref{secondorder}) with the linear solutions $h^{(1)}_{ab}$ found by solving Eq.(\ref{linear}) etc.  
$G^{(1)}$ is a linear differential operator. The operator $G^{(2)}[h]$ produces 24 terms which are schematically either type $h\partial^2 h$ or type $(\partial h)^2$ with different index contractions. 
%
%
Then, one computes $G^{(2)}[h^{(1)}]$ by inserting the obtained solutions. Therefore $G^{(2)}[h^{(1)}]$ is a known function and then equation (\ref{secondorder}) is just a set of linear differential equations for $h^{(2)}$. The solution will seed the third order perturbation equations, and so on. In this way, one can find approximate solutions to Einstein equations by solving iteratively, a set of {\it linear} differential equations.

\section{Solutions of perturbative gravity}\label{solgrav}

Starting with Minkowski space, one can find perturbative solutions of pure gravity. In this section we will review the  first and second order equations and solutions in the Lorentz gauge, 
since these will be compared to their bilocal perturbative counterparts. 
The reader can find a complete analysis of these solutions in \cite{APR}.

First order equations of Einstein gravity are well-known and lead to the gravitational wave equations in an appropriate gauge, see for instance section 4.4 of \cite{RW}. Linearized gravity equations (\ref{linear}) are of the form
\begin{equation}\label{lineargen}
-\Box h_{\mu\nu}+h^{\phantom{1}\alpha}_{\nu \phantom{1},\mu\alpha}+h^{\phantom{1}\alpha}_{\mu \phantom{1},\nu\alpha}-h_{,\mu\nu}-h^{\alpha\beta}_{\phantom{1},\alpha\beta}\eta_{\mu\nu}+\eta_{\mu\nu}\Box h=16\pi G T_{\mu\nu}.
\end{equation}
Note that in this section we will be writing $h_{\mu\nu}$ instead of $h^{(1)}_{\mu\nu}$ in order to simplify notation.
Since we consider only pure gravity, $T_{\mu\nu}=0$. 
Upon taking {\it trace-reversed} variables
\\
\begin{equation*}
\bar{h}_{\mu\nu}=h_{\mu\nu}-\frac{1}{2}h\eta_{\mu\nu},
\label{hbar}
\end{equation*}
equation (\ref{lineargen}) reads:
\begin{equation}\label{tracereversedeq}
-\Box \bar{h}_{\mu\nu}+\bar{h}^{\phantom{1}\alpha}_{\nu \phantom{1},\mu\alpha}+\bar{h}^{\phantom{1}\alpha}_{\mu \phantom{1},\nu\alpha}-\bar{h}^{\alpha\beta}_{\phantom{1},\alpha\beta}\eta_{\mu\nu}=0~.
\end{equation}
Next, imposing the Lorentz gauge condition \\
\begin{equation}\label{lorentz}
\partial^{\mu}\bar{h}^{(1)}_{\mu\nu}=0,
\end{equation}
reduces (\ref{tracereversedeq}) to the standard wave equation
\begin{equation}\label{box}
\Box \bar{h}^{(1)}_{\mu\nu}=0~.
\end{equation}
Eq.(\ref{lorentz}) does not fix the gauge completely. The reason is that we can always gauge transform $h_{\mu\nu}\to h_{\mu\nu}+\xi_{\mu,\nu}+\xi_{\nu,\mu}$ with $\Box\xi_{\mu}=0$ and (\ref{lorentz}) still holds \cite{RW}.  \\
Further, imposing the transverse-traceless gauge condition it can be shown that the general solution of (\ref{box}) can be written as
\begin{equation}\label{hbarlinear}
\bar{h}^{(1)}_{\mu\nu}=\Re \big(B_{\mu\nu}e^{iPx}\big),
\end{equation}
where $P_\mu$ is a null vector,  and
\begin{equation}\label{Alinear}
\mathbf{B} = \left(
\begin{array}{cccc}
0 & 0 & 0&0 \\
0& B_+ & B_{\times}&0 \\
0& B_{\times}&-B_+&0 \\
0&0&0&0
\end{array} \right).
\end{equation}
After complete gauge-fixing, 
The two remaining degrees of freedom, $B_+$ and $B_{\times}$ corresponding to the 
two polarizations, are true physical degrees of freedom.

Although the second order equations (\ref{secondorder}) are quite involved, they  
once again simplify greatly on gauge fixing. One treats the 
first order solutions (\ref{hbarlinear}) with (\ref{Alinear}),
as sources for the second order equations. Then in the 
trace-reversed coordinates and the Lorentz gauge at second order 
\begin{equation}
\partial^{\mu}\bar{h}^{(2)}_{\mu\nu}=0,
\end{equation}
we arrive at the equations
\begin{equation}\label{graveq}
\Box \bar{h}^{(2)}_{\mu\nu}=-\frac{1}{2}\text{Tr}({\bf B}^2) P_{\mu}P_{\nu}\text{ exp}[i2Px],
\end{equation}
where ${\bf B}$ is the polarization matrix (\ref{Alinear}). 

It can be shown \cite{APR} that the monochromatic wave solutions of (\ref{graveq}) are 
\begin{equation}
\bar{h}^{(2)}_{\mu\nu}=\Re\Big[\Big(a_{\mu\nu}+ib_{\mu\nu}\frac{J_{\sigma}x^{\sigma}}{J_{\sigma}P^{\sigma}}\Big)\text{ exp}[i2Px]\Big],
\end{equation}
where $J_{\sigma}$ is an arbitrary four-vector, and 
\begin{eqnarray}\label{abgravity}
a_{\mu\nu}&=&-\frac{1}{16}\text{Tr}({\bf B}^2)\eta_{\mu\nu}\nonumber \\
b_{\mu\nu}&=&\frac{1}{8}\text{Tr}({\bf B}^2) P_{\mu}P_{\nu}.
\end{eqnarray}

\section{Bilocal model}\label{BM}

Next, we turn to the bilocal fields. The 
bilocal model we consider here
is similar to those proposed in \cite{Yu,T,bilocal,Fe} for the description of hadrons. 
However, the important 
difference is that our model is adapted to accommodate massless solutions, 
which were not studied before. As we shall see, 
this is nevertheless crucial for the connection with gravity.

\subsection{Bilocal field}

The basic object of a bilocal model is the bilocal field
\begin{equation*}
\bar{\Phi}(x,y),
\end{equation*} 
which as indicated, depends on two points of spacetime as opposed to a single point as in local theories. 
It is useful to work in the `more physical' 
{\it center of mass} (CM) and {\it relative} coordinates. The former is defined as:
\begin{equation*}\label{coor}
X_{\mu}=\frac{1}{2}(x_{\mu}+y_{\mu}), 
\end{equation*}
and are the ones that are observable at large distances and survive in the local limit. 

The  {\it relative} or {\it internal} coordinates on the other hand 
represent the distance between $x$ and $y$. Here we define the Euclidean version
of this coordinate as follows:
\begin{equation}\label{relativecoor}
 r_{\mu}=(x_{\mu}-y_{\mu})_E,\quad r^2=r^{\mu}r^{\nu}\eta_{\mu\nu}\geq 0~.
\end{equation}
%
This avoids physical inconsistencies 
associated with having `two times', as well as the violation of causality,
which would accompany action-at-a-distance in Lorentzian internal coordinates. \\
Also note that issues related to causality may indeed arise if non-local effects are observed at low energies.
However in the solutions of our bilocal model, there is a Gaussian term $e^{-\frac{\alpha^2}{2}r^2}$ which suppresses nonlocal effects for positive $r^2$,
which is indeed the case for Euclidean relative coordinates. 
A Minkowskian relative coordinate on the other hand would have a growing time mode leading to unphysical solutions. 
The Wick rotation in the relative coordinate cures this problem. Such a rotation can also be seen as a canonical transformation of the initial variables $(x^{\mu}+y^{\mu},x^{\mu}-y^{\mu})$ which leads to an equal-time constraint; details of this procedure can be found 
in \cite{Kam} for the case of massive solutions and \cite{JJY} for a general treatment which includes the massless sector
\footnote{ A general discussion of how to obtain a theory with one time coordinate out of a theory with two times is contained in   
in \cite{Bar} and references therein.
Euclidean internal coordinates for the bilocal field were also used in the context of AdS/CFT  
\cite{jevicki1}, although in that paper the CM coordinates were also Euclidianized for an Euclidean path integral formulation.}.

We now write the bilocal field in terms of these new coordinates:
\begin{equation*}
\Phi(X,r)=\bar{\Phi}(x,y).
\end{equation*}
%
As is customary, 
we will consider the bilocal field symmetric under the exchange $x\leftrightarrow y$. 
That is, $\bar{\Phi}(x,y)=\bar{\Phi}(y,x)$ or equivalently, 
\begin{equation}\label{evencond}
\Phi(X,r)=\Phi(X,-r).
\end{equation}
This ensures that neither of the internal points $x$ or $y$ is preferred, and that 
physics is invariant under their interchange. 
Note that this symmetry is not always imposed for bilocal fields. 
In the context of vector models, the symmetry is usually absent whereas for fermion fields (in the SYK context) 
the field is antisymmetric under the exchange of $x$ and $y$. 
In addition to the perturbative expansion (\ref{phiexp}), 
the bilocal field functions may now be expanded around locality as
\begin{equation}\label{Taylor2}
\Phi^{(i)}(X,r)=\phi(X)^{(i)}+H_{\mu\nu}^{(i)}r^{\mu}r^{\nu}+D_{\mu\nu\sigma\rho}^{(i)}r^{\mu}r^{\nu}r^{\sigma}r^{\rho}\cdots,
\end{equation}  
where
\begin{eqnarray}\label{fieldtower}
\phi^{(i)}(X)&\equiv&\Phi^{(i)}(X,0) \nonumber \\
H_{\mu\nu}^{(i)}(X)&\equiv&\frac{1}{2}\partial_{\mu}\partial_{\nu}\Phi^{(i)}(X,r)\Big|_{r=0} \nonumber \\
D_{\mu\nu\sigma\rho}^{(i)}(X)&\equiv&\frac{1}{4!}\partial_{\mu}\partial_{\nu}\partial_{\sigma}\partial_{\rho}\Phi^{(i)}(X,r)\Big|_{r=0} \nonumber \\
&\vdots&
\end{eqnarray}
We see that the effect of the bilocal field is to bring in
new degrees of freedom, namely the dependence of the field on the relative coordinates. They result in an infinite tower of high (even) spin fields $H_{\mu\nu}^{(i)}(X),D_{\mu\nu\sigma\rho}^{(i)}(X)\dots$, as seen in (\ref{fieldtower}). 
We will show in Section \ref{GfB}
that, at least up to second order, gravity is encoded in these extra degrees of freedom. 
We would like to clarify that those fields do not correspond one-to-one to the {\it physical} gravity and higher spin fields. As already noticed in the earlier paper \cite{jevicki1}, the relation seems non-local. Specifically, the physical gravity field will not be identified with $H_{\mu\nu}$ but with a suitable linear combination of the field expansion. In this work, we work out the specific relation between the bilocal field and gravity at the linear and second order in the perturbative series. We do not provide a general prescription for the identification at higher orders although we give enough evidence of its plausibility.  \\
We mention here that since self-interacting terms of the bilocal field induce interactions in the higher spin fields, and that theories 
describing such fields in flat spacetimes (namely the Fronsdal's program) present  
apparent inconsistencies beyond cubic vertices, one may be concerned thats similar inconsistencies may also arise here
\footnote{See \cite{BBSu} for a discussion on the consistency of Fronsdal's approach.}.  
%
%
However, note that a crucial assumption in those theories is locality, which restricts the number of derivatives of the fields in the Lagrangian, 
and dropping locality leads to a possible breakdown of causality.
On the other hand in our case, the locality assumption should not be considered, since  
micro-causality is already violated by the nonlocal nature of the bilocal field. 
Therefore, the aforementioned inconsistencies do not arise in our case. 
Nonetheless, we believe that this is an important point worth investigating further.

\subsection{Bilocal equations}

The classical bilocal scalar field model we are considering (similar to other models \cite{T,bilocal,Fe}) consists of two equations:
\begin{eqnarray}
\big(\Box+\Box_r-\alpha^4r^2+8\alpha^2\big)\Phi(X,r)&=&V(\Phi)\label{eom}\\
\frac{\partial}{\partial X_{\mu}}\bigg(\frac{\partial}{\partial r^{\mu}}+\alpha^2 r_{\mu}\bigg)\Phi(X,r)&=&0 \label{constraint},
\end{eqnarray}
where $\Box$ and $\Box_r$ are the d'Alembertians associated with coordinates $X_{\mu}$
and $r_{\mu}$ respectively. The parameter $\alpha$ has dimensions of mass and controls bilocality, in the sense that locality is recovered as $\alpha\to \infty$. Equations (\ref{eom}) and (\ref{constraint}) could be derived from an action, as done for a similar (although non-interacting) model in \cite{T3}. A bilocal Lagrangian will be necessary, in a future work, when comparing the bilocal theory and gravity at the level of actions. In this paper we explore and connect the respective spaces of solutions, for which we just need the equations of motion.   

\begin{figure}
\centering
  \includegraphics[scale=0.37]{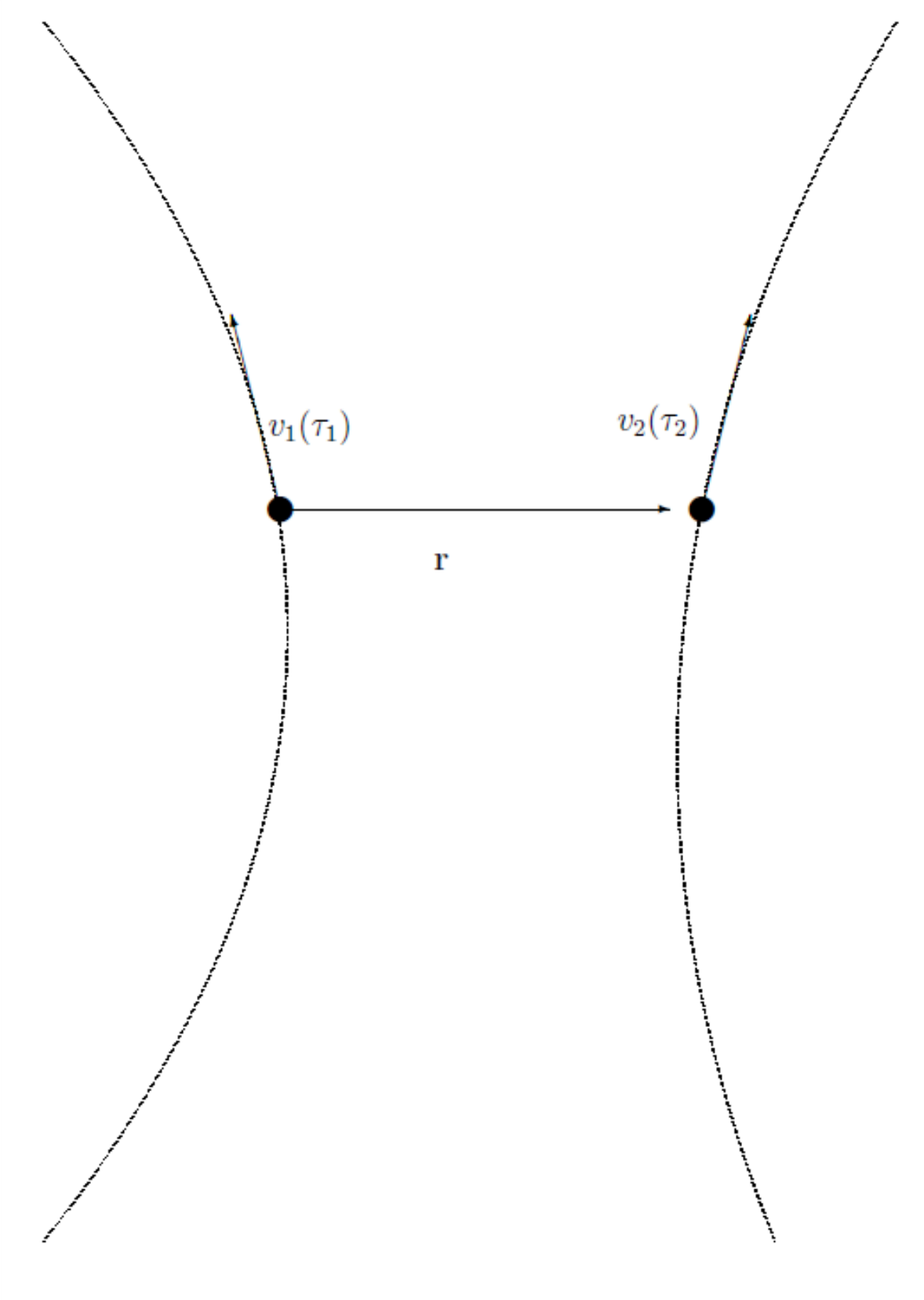}
  \caption{In the point particle picture, the bilocal equations describes the motion of a bound system of two particles. The equations (\ref{eom}) and (\ref{constraint}) translate into a harmonic instant interaction between the two particles.}
  \label{pointparticle}
\end{figure}
%

Note that Eq.(\ref{eom}) is dynamical whereas Eq.(\ref{constraint}) is a constraint, as can be guessed at first sight since eq. (\ref{constraint}) only involves  first derivatives of the CM coordinates.
We can see from the equations (\ref{eom}) and (\ref{constraint}) that the model has the minimal ingredients to be short-ranged as well as genuinely bilocal. 
The term  $\alpha^4r^2$ in Eq.(\ref{eom}) forces short-ranged non-locality, since it makes the general solutions of the model have a Gaussian decay factor in the relative coordinates.

It is easier to understand the implications of these two equations if we shift
for a moment from the field theory to a point-particle description (Fig.\ref{pointparticle}). In this figure, we see the worldlines of the two-particle system that the model describes. Parameters $\tau_1$ and $\tau_2$ are their respective proper times. Now, equation (\ref{constraint}) is derived from the simple kinematical relation ${\bf r}\cdot({\bf v}(\tau_1)+{\bf v}(\tau_2))=0$, which is independent of the reparametrizations of $\tau_1$ and $\tau_2$. 
This geometrical constraint implies a condition $\tau_1(\tau_2)$, 
which in turn implies an action-at-a-distance between the two particles \cite{T2}. In summary, Eq.(\ref{constraint}) serves to identify the proper time corresponding
to the points $x$ and $y$, which in turn ensures an action-at-a-distance.
Note that different constraints (and so different actions-at-a-distance) could have been chosen \footnote{See \cite{T2}, for a comprehensive analysis of these constraints. When comparing with gravity, the constraint equation translates into the Lorentz gauge. It seems reasonable to associate the arbitrariness of the choice of constraint with the freedom of choosing different gauges in gravity.}. 
Eq.(\ref{eom}) and Fig.\ref{pointparticle} show
that the particles interact as if linked by a spring, with the spring
constant controlled by parameter $\alpha$.   \\

It is generally accepted that gravitons are massless. Thus, we will study the general massless solutions of the model (\ref{eom}) and (\ref{constraint}) in the next section, and we will show how they naturally encode gravitational waves at linear order and reproduce the second order gravity solutions as well. 

\subsubsection{Solutions to linear order}
\label{LO}

To first (linear) order, we have from Eqs.(\ref{eom}) and (\ref{constraint}), 
\begin{eqnarray}
\big(\Box+\Box_r-\alpha^4r^2+8\alpha^2\big)\Phi^{(1)}(X,r)&=&0\label{linmodel}, \\
\frac{\partial}{\partial X^{\mu}}\bigg(\frac{\partial}{\partial r^{\mu}}+\alpha^2 r_{\mu}\bigg)\Phi^{(1)}(X,r)&=&0\label{linconstraint}.
\end{eqnarray}

Note that the model is slightly different\footnote{Terms in (\ref{linconstraint}) are summed instead of subtracted. This modification is irrelevant in \cite{DDW} but it is necessary in the present context in order to keep the solutions short-ranged with the spacetime signature $(-,+,+,+)$. The addition of the constant term $8\alpha^2$ is necessary to keep the general solutions as (\ref{linansatz}), well adapted to seed higher order contributions. Note that the factor 8 would change into $d+4$ in a $d$-dimensional spacetime.} to the one we used in \cite{DDW}. The small changes in the equations, that translate into the shape of the solutions (the matrix $\bar{C}_{\mu\nu}$ which carries the internal degrees of freedom no longer appears in the exponential, see 
Eq.(\ref{linansatz}) below), are chosen for the model to be adapted to higher orders in perturbation theory. In other words, the model used in \cite{DDW} works perfectly fine for linear order, whereas the slightly modified model we propose here is valid for the comparison with gravity at  linear {\it and} higher orders. The existence of a bilocal model adapted to describing gravity at linear order \cite{DDW}, whereas not suitable for higher orders may reflect the especial role of linearized gravity as a complete theory in its own. 

Let us take the ansatz 
\begin{equation}\label{linansatz}
\Phi^{(1)}(X,r)=r^{\mu}\bar{C}_{\mu\nu}r^{\nu}e^{-\frac{\alpha^2}{2}r^2}e^{iPX},
\end{equation}
where $\bar{C}_{\mu\nu}$ is an yet undetermined  $4\times 4$ matrix. The momentum $P_{\mu}$ is a 4-vector with dimension of mass. \\
Substituting (\ref{linansatz}) into (\ref{linmodel}) leads to 
\begin{equation}
2\eta^{\mu\nu}\bar{C}_{\mu\nu}-P^2r^{\mu}r^{\nu}\bar{C}_{\mu\nu}=0
\end{equation}
and substituting (\ref{linansatz}) into (\ref{linconstraint}) leads to
\begin{equation}
P^{\mu}\bar{C}_{\mu\nu}r^{\mu}=0.
\end{equation}

Now, since the above conditions should hold for all $r^{\mu}$ we see that equations
(\ref{linmodel}) and (\ref{linconstraint}) imply

%
\begin{eqnarray}
\eta^{\mu\nu}\bar{C}_{\mu\nu} &=& 0 , \label{A1} \\
 P^{\mu}P_{\mu}&=&0,\quad\text{(massless)} \\
 P^{\mu}\bar{C}_{\mu\nu} &=& 0. \label{A2}
\end{eqnarray}
From Eq.(\ref{A1}) and the symmetric nature of $\eta^{\mu \nu}$ it follows that  
$\bar{C}_{\mu\nu}$ is symmetric as well. 
Then it follows from Eqs.(\ref{A1}) and (\ref{A2}), 
that the general solution (\ref{linansatz}) is of the form
\begin{equation}\label{firstorderrelations}
\bar{C}_{\mu\nu}=C_{\mu\nu}+c P_{\mu\nu},\quad P_{\mu\nu}\equiv \frac{P_{\mu}P_{\nu}}{p^2},\quad (p^2=P_0^2),
\end{equation}
where $C_{\mu\nu}$ is a symmetric matrix. The matrix elements $P_{\mu\nu}$ are defined in (\ref{firstorderrelations}) to be dimensionless. From eq. (\ref{A1}), we see that all solutions with shape (\ref{linansatz}) are massless, a condition that will extend to the higher spin fields. Massive solutions would require a modification of (\ref{linmodel}), namely changing the 8 factor in $8\alpha^2$.

Next, without loss of generality, we choose a set of coordinates such that direction of propagation of the CM, $P_{\mu}$, is parallel to the $z$-axis. In these coordinates we have the general solution
\begin{equation}\label{goodsols}
{\bf \bar{C}}=\left(
\begin{array}{cccc}
c& 0 & 0 &c\\
0&a & b&0\\
0&b&-a&0\\
c& 0 & 0 &c
\end{array} \right),
\quad {\bf C}=\left(
\begin{array}{cccc}
0& 0 & 0 &0\\
0&a & b&0\\
0&b&-a&0\\
0& 0 & 0 &0
\end{array} \right),
\end{equation}
where $c,a,b$ are real constants. 
From the ansatz (\ref{linansatz}) we see that the parameter $\alpha$  measures non-locality. The limit
\begin{equation*}
\lim_{\alpha\to \infty}\frac{\alpha}{\sqrt {2\pi}}e^{-\frac{1}{2}\alpha^2 x^2}=\delta(x),
\end{equation*} 
applied to each coordinate, implies that the relative dimensions vanish for large $\alpha$. So,  in the limit $\alpha\to \infty$ the space of relative coordinates shrinks and one is left with a  local theory.
We see that the parameter $\alpha$ is a measure of locality in the model, and that the theory is more local with increasing $\alpha$. 
Also the perturbative expansion driven by $K_B$, is performed around locality. 
Then since $\alpha$ is the only parameter of the theory, $K_B$ must be related to it by $K_B=(1/\alpha)^m$ 
for some positive $m$. One thus expects the perturbative expansion in the bilocal model to work well for large $\alpha$.

\subsubsection{Solutions to second order}\label{SO}

The most natural second order term for $V(\Phi)$ in bilocal theories is of the form 
\begin{equation}\label{V2}
V^{(2)}(x,y)\sim\int\Phi^{(1)}(x,z)\Phi^{(1)}(y,z).
\end{equation}
This is because the interaction term must be bilocal and must involve two copies of $\Phi^{(1)}$. From (\ref{V2}), it is straightforward to see that $V^{(2)}(x,y)=V^{(2)}(y,x)$, as it should be for consistency. \\
Explicitly, the bilocal equations to second order we consider are
\begin{eqnarray}
\big(\Box+\Box_r-\alpha^4r^2+8\alpha^2\big)\Phi^{(2)}(X,r)&=&e^{\frac{\alpha^2}{2}r^2}\int \bar{\Phi}^{(1)}(x',z) \bar{\Phi}^{(1)}(z,y') dz,\label{secondmodel}\\
\frac{\partial}{\partial X^{\mu}}\bigg(\frac{\partial}{\partial r^{\mu}}+\alpha^2 r_{\mu}\bigg)\Phi^{(2)}(X,r)&=&0
\label{secondconstraint}
\end{eqnarray} 
where $\bar{\Phi}^{(1)}(x,z)\equiv \Phi^{(1)}(X,r)$ 
and $\Phi^{(1)}(X,r)$ are the solutions of the first order equations (\ref{linansatz}) with (\ref{goodsols}). 
The points $x',y'$ of (\ref{secondmodel}) are defined as 
\begin{equation}
x'+y'=x+y,\quad x'-y'=2(x-y),
\end{equation}
which gives 
\begin{equation}\label{primecoordinates}
X'=X, \quad r'=2r.
\end{equation}
The specific form of the RHS of (\ref{secondmodel}) is explained as follows. The prefactor $e^{\frac{\alpha^2}{2}r^2}$ serves to force $\Phi^{(2)}(X,r)\propto e^{-\frac{1}{2}\alpha^2 r^2}$, as  $\Phi^{(1)}(X,r)$ does. So the global solution, $\Phi=\Phi^{(1)}+\kappa_B\Phi^{(2)}+\cdots$ has an overall Gaussian decay $e^{-\frac{1}{2}\alpha^2 r^2}$. This condition for the global solution, which does not seem essential but simplifies the treatment when compared to gravity, could apply to every higher order interacting terms resulting in similar pre-factors
\footnote{In the absence of the prefactor, one has $\Phi^{(2)}(X,r)\propto e^{-\alpha^2 r^2}$, 
and subsequent orders in the perturbative expansion would have different (sharper) Gaussians. 
As mentioned above, since this does not affect the treatment in an essential way, 
we will consider dropping this factor in future works.}. 
The change of coordinates (\ref{primecoordinates}) is necessary to assure that the the second order solutions are $\Phi^{(2)}(X,r)\propto P^{(4)}(r)e^{\frac{-\alpha^2}{2}r^2}$, as we will see later. As shown in section \ref{GfB}, the details of the polynomial $P^{(4)}(r)$ will eventually encode $\bar{h}^{(2)}_{\mu\nu}$. The same happens at first order where $\Phi^{(1)}(X,r)\propto P^{(2)}(r)e^{\frac{-\alpha^2}{2}r^2}$, and $P^{(2)}(r)$ encodes gravitational waves. In general, we propose to constraint the higher order interaction terms so that $\Phi^{(n)}(X,r)\propto P^{(2n)}(r)e^{\frac{-\alpha^2}{2}r^2}$, and then the contribution $\bar{h}^{(n)}_{\mu\nu}$ can be read from $P^{(2n)}(r)$. This will result into an essentially unique shape for $V(\phi)$ which could encode the exact gravity self-interacting terms. \\
One
can compute the integral that appears on the RHS of (\ref{secondmodel}), 
we do it in Appendix \ref{Gauss}. In that calculation
we see that the integral splits as
\begin{equation}
I'(X,r)=\int \bar{\Phi}^{(1)}(x',z) \bar{\Phi}^{(1)}(z,y') dz=I'(r) e^{i2PX}.
\end{equation}
Next, we plug in the ansatz
\begin{equation}
\label{ansatz2order}
\Phi^{(2)}(X,r)=\big[A(r)+\frac{1}{2i}B^{\sigma}(r)X_{\sigma}\big]e^{i2PX},
\end{equation}
with $A(r)$ and $B^{\sigma}(r)$ real, into (\ref{secondmodel}). Notice that the 4-vector $P_{\mu}$ is the same as introduced for the linear solutions. So $P^2=0$ and the second order contribution is also massless\footnote{The same will happen for higher order contributions, which are plane waves $e^{i3PX},e^{i4PX},\dots$ in the CM coordinates.}.  It
results in 
%
\begin{eqnarray}
&& e^{\frac{\alpha^2}{2}r^2}I'(r)=
\big(\Box_r-\alpha^4r^2+8\alpha^2\big)A(r)+B(r),\quad B(r)\equiv P_{\sigma} B^{\sigma}(r) \label{ABmodel}\\
&&\big(\Box_r-\alpha^4r^2+8\alpha^2\big)B^{\sigma}(r) = 0\label{Bmodel}
\end{eqnarray}
while plugging in the ansatz (\ref{ansatz2order}) into Eq.(\ref{secondconstraint}) yields 
\begin{eqnarray}
&&(\partial_{\mu}+\alpha^2r_{\mu}) B^{\mu}(r)-4P^{\mu}(\partial_{\mu}+\alpha^2r_{\mu})A(r)
=0 , \label{ABconstraint} \\
&& P^{\mu}(\partial_{\mu}+\alpha^2r_{\mu})B^{\sigma}(r) =0 . \label{Bconstraint}
\end{eqnarray}
Since Eqs.(\ref{Bmodel}) and (\ref{Bconstraint}) involve only 
the function $B^\sigma(r)$, we solve them first. It can be shown that the function
\begin{equation}\label{bsigma}
B^{\sigma}(r)=e^{-\frac{\alpha^2}{2}r^2}\big(B^{\sigma}_0P_{\mu\nu}+B^{\sigma}_1C_{\mu\nu}\big)r^{\mu}r^{\nu},
\end{equation}
where $B^{\sigma}_0$ and $B^{\sigma}_1$  are constant vectors, solves both equations. 
There is an abuse of notation in eq. (\ref{bsigma}) since at this stage, $C_{\mu\nu}$ need not be the same matrix as the one at linear order. 
It will need to coincide later though, as we use the other two equations. 
Now, since Eqs.(\ref{Bmodel}) and (\ref{Bconstraint}) are identical to (\ref{linmodel}) and (\ref{linconstraint}) for the $r$ dependence, 
the constraints (\ref{A1}) and (\ref{A2}) on the solutions are the same. 
Particularly $P^{\mu}C_{\mu\nu}=0$, which fact we have used in (\ref{bsigma}). \\

Next, let us find the function $A(r)$. Plugging in
$B^{\sigma}(r)$ from Eq.(\ref{bsigma}) in (\ref{ABconstraint}), we get 
\begin{equation}\label{eqA1}
P^{\mu}(\partial_{\mu}+\alpha^2r_{\mu})A(r)=\frac{B_0}{2p^2}~(P_{\mu}r^{\mu})e^{-\frac{\alpha^2}{2}r^2},\quad B_0\equiv P_{\mu}B^{\mu}_0,
\end{equation}
where we have chosen\footnote{This restricts the space of solutions. We do not mind at this stage since we are not looking for the most general second order bilocal contribution. It is enough for our purposes to find the set of bilocal solutions that encodes the most general gravity solution of the same order. } $B_1^{\sigma}$ so that $B_1^{\sigma}C_{\sigma\nu}=0$. \\
We assume the form 
\begin{equation}\label{Aansatz}
A(r)=A_0r^2e^{-\frac{\alpha^2}{2}r^2}+\dots
\end{equation}
where the dots are terms that vanish under the action of $P^{\mu}(\partial_{\mu}+\alpha^2r_{\mu})$ and that will be fixed later. 
Substituting this into Eq.(\ref{eqA1}), one arrives at the condition 
\begin{equation}
A_0=\frac{B_0}{4p^2},
\end{equation}
remind that $p^2=P_0^2.$

Let us find the functions $A(r)$ and $B(r)$.
In order to determine $A(r)$ we must solve (\ref{ABmodel}) and use the precise function $I'(r)$ shown in (\ref{integral}).
To simplify notation, let us call 
\begin{equation}
K=\frac{\pi^2}{\alpha^8}e^{-\frac{p^2}{2\alpha^2}}.
\end{equation}

Comparing terms in (\ref{ABmodel}), functions (\ref{bsigma}) get fixed as
\begin{equation}\label{finalB}
B(r)=P_{\sigma}B^{\sigma}(r)=2K(\alpha^2-p^2)\bigg(c C_{\mu\nu}-c^2 P_{\mu\nu}\bigg)r^{\mu}r^{\nu}e^{-\frac{\alpha^2}{2}r^2},
\end{equation}
and
\begin{eqnarray}\label{finalA}
&&A(r)=e^{-\frac{\alpha^2}{2}r^2}K\bigg[J (Pr)^2+\frac{c^2}{2}(\alpha^2-p^2)r^2-\frac{1}{4}\alpha^2(C_{\mu\nu}r^{\mu}r^{\nu})^2 
-\frac{c^2}{4}\alpha^2(Pr)^4\nonumber \\
&&-\frac{c}{2}\alpha^2 C_{\mu\nu}r^{\mu}r^{\nu}(Pr)^2+\frac{c^2}{4\alpha^2}\Big(\frac{p^4}{\alpha^4}-6\frac{p^2}{\alpha^2}+3-2(\alpha^2-p^2)\Big)+\frac{1}{8\alpha^2}\text{Tr}(C^2)\Big)\bigg], \nonumber \\
\end{eqnarray}
where $J$ is an arbitrary constant. 
Therefore, functions (\ref{finalB}) and (\ref{finalA}),
together with (\ref{ansatz2order}) solve the second order equations 
(\ref{secondmodel}) and (\ref{secondconstraint}) of the model. \\

\section{Gravity from bilocality}
\label{GfB}

Next we show that one
can recover gravity solutions as a subset of the bilocal solutions we found in the last section.
First, at linear order, note that gravity solutions in the Lorentz 
transverse traceless gauge (gravitational waves) are obtained from the bilocal general linear solutions, (\ref{linansatz}) with (\ref{goodsols}), if we make $c=0$. 
As mentioned in the introduction, the bilocal model we are considering which has no gauge symmetry, corresponds to gravity in the Lorentz gauge. 
However, the residual gauge of gravity seems not to be fixed by the model. 
Taking the particular subset of the solutions $c=0$ for the bilocal fields corresponds to fixing the residual gauge in the gravity side.\\
Specifically, we have 
\begin{equation}
\bar{h}^{(1)}_{\mu\nu}(X)=\Re \big(\partial_{\mu}\partial_{\nu}\Phi^{(1)}(X,r,c=0)\big|_{r=0}
\big)=\Re\big(C_{\mu\nu}e^{iPX}\big),
\end{equation}
where we have identified the center of mass coordinates of the bilocal field with 
the ordinary spacetime in gravity.

The set of bilocal solutions gets considerably reduced at second order with $c=0$. The Gaussian integral (\ref{gaussianintegral}) is 
\begin{equation}
I(X,r,c=0)=e^{i2PX}e^{\frac{-\alpha^2}{4}r^2}\frac{K}{2}\bigg[
\text{Tr}(C^2)-\alpha^2C^2_{\mu\nu}r^{\mu}r^{\nu}
+\frac{1}{8}\alpha^4(C_{\mu\nu}r^{\mu}r^{\nu})^2\bigg],
\end{equation}
and 
\begin{equation}
B^{\sigma}(r,c=0)=0.
\end{equation}
So, for $c=0$, the second order solutions (\ref{ansatz2order}) simplifies to 
\begin{equation}
\Phi^{(2)}(X,r,c=0)=A(r,c=0)e^{i2PX},
\end{equation}
with (\ref{finalA}) turning into
\begin{equation}\label{Amzero}
A(r,c=0)=e^{-\frac{\alpha^2}{2}r^2}K\bigg[J (Pr)^2-\frac{1}{4}\alpha^2(C_{\mu\nu}r^{\mu}r^{\nu})^2 
+\frac{1}{8\alpha^2}\text{Tr}(C^2)\Big)\bigg].
\end{equation}
The gravity second order contributions sourced by gravitational waves are reproduced in the bilocal picture upon the identifications
\begin{eqnarray}\label{secondorderrelations}
a_{\mu\nu}&=&-\frac{1}{2}\frac{\alpha^2}{K}A\big(0,c=0\big)\eta_{\mu\nu},\nonumber \\
 b_{\mu\nu}&=&\frac{\alpha^2}{K}\Big(\partial_{\mu}\partial_{\nu}A^2(r,c=0)\big|_{r=0}+\alpha^2A^2(r,c=0)\eta_{\mu\nu}
\Big)\nonumber \\
B_{\mu\nu}&\equiv &C_{\mu\nu}
\end{eqnarray}
Note that the quantities on the LHS of (\ref{secondorderrelations}) determine a general  solution of gravity up to second order, see (\ref{abgravity}).
Relations (\ref{secondorderrelations}) together with (\ref{firstorderrelations}) for $c=0$ are the main results of the paper.
They show that gravity solutions, at least up to the second order contribution, are encoded in a subset of the bilocal solutions. This is a non-trivial connection and it is likely to hold for subsequent orders in perturbation theory.

It is surprising that both theories match given that self-interactions in gravity are driven by terms schematically of the form $\partial \partial h^n$ whereas bilocal counterparts are $\int \Phi^n$. Technically, this is possible because of a peculiar behaviour of multivariable integrals of Gaussians.
In general, for any invertible $d \times d$-matrix $M$ and a polynomial $P(r)$ on the variables $r_1,\dots,r_d$, we have
\begin{equation}\label{mathidentity}
\int P(r)e^{-\frac{1}{2}M_{\mu\nu}r^{\mu}r^{\nu}}\text{d}^dr=\sqrt{\frac{(2\pi)^d}{\text{det}M}}e^{\frac{1}{2}(M^{-1})_{\mu\nu}\partial^{\mu}\partial^{\nu}}P(r)\Big|_{r=0},
\end{equation}
where the exponential of the derivatives are understood as a power series. This identity can be proved by an iterative integration by parts. Notice that all order contributions $\Phi^{(i)}$ will be of the form $P^{(2i)}(r)e^{-\frac{1}{2}\alpha^2r^2}$. Thus, a general interaction term will always involve integrals of polynomials with Gaussians.
As a physical consequence, the mathematical identity (\ref{mathidentity}) is behind the connection between bilocal models and gravity we are claiming.

\section{Connection to holography} 
\label{CWH}
The connection between bilocal theories and gravity makes it possible to study gravity (and quantum gravity once the bilocal theory is quantized) from a theory which does not {\it a priori} have gravitational degrees of freedom. 
This resembles holography, where gravity in $d+1$ dimensions is recovered from a (conformal) non-gravitational theory in $d$ dimensions. 
We will sketch how this connection works
\footnote{We leave a detailed analysis for a future work.}, but before that, 
let us point out some obvious differences between the two pictures.
First, the bilocal model is non-local whereas both sides in holography 
can be described by local theories, at least in the regime where true stringy effects can be ignored, and one 
can consider perturbative Einstein gravity on the gravity side, and a strongly coupled CFT with large $N$ $U(N)$ gauge group on the gauge theory side (for AdS/CFT)
\footnote{Non-local effects are natural in the AdS/CFT correspondence in regimes where stringy effects in the gravity side are relevant. 
However holography in its entirety implies non local effects. In another example, higher spin theories in AdS (Vasiliev's theory) which are supposed 
to be dual to vector CFT's, invariably involve non-local interactions. We thank the Referee for pointing this out.}.
Second, in the bilocal model, both the bilocal field theory and gravity are in $d+1$ dimensions, 
in contrast with holography, where the gauge theory is $d$-dimensional. 
Third, the bilocal theory is manifestly non-conformal, 
since it involves the dimensional parameter $\alpha$, which measures the degree of non-locality.
One advantage of our approach is that the `bulk spacetime' need not be asymptotically AdS;
in fact it can be asymptotically flat. 
\\
However, there is a double limit that can lead to a compatibility between the two pictures.   
 Let us consider
\begin{equation}\label{limit}
\lim_{\substack{\alpha\to \infty\\
r\to 0}}\alpha r=\lambda,
\end{equation}  
where $r$ is the magnitude of the relative coordinate. In this picture it is convenient to decompose the relative coordinate into its magnitude and angles, and the latter is used to expand the bilocal field in spherical harmonics \cite{jevicki1}, which are considered internal degrees of freedom. In this way, the bilocal field emerges as a theory in $d+1$ dimensions, the extra dimension being $r$. This very coordinate, $r$, was found to be proportional to the radial coordinate of $AdS_4$ in the most precise reconstruction of $AdS$ from the collective (bilocal) field in the CFT \cite{dMJJR}. 

The parameter $\lambda$ is dimensionless by construction, and could be interpreted as the coupling constant of the theory that emerges at the limit. The limit theory is defined at $r=0$, so it is $d$-dimensional. It is local, since $\alpha\to \infty$, and it is potentially conformal since the coupling is dimensionless. It seems that the  connection between the bilocal picture  and holography must be through the limit (\ref{limit}), which is tantamount to sticking to the boundary theory in the context of holography. The bilocal construction and holography seem perfectly compatible. Moreover, it is likely that starting from the bilocal field as a fundamental object one could derive naturally holographic relations, a path that would be interesting to explore in the future.   

The reader may wonder about the singular behaviour of the solutions like (\ref{linansatz}) in the double limit (\ref{limit}). 
It may seem that the solution would shrink to $0$ in that limit. 
However, note that solutions are not normalized. The dimension of the bilocal field (which some authors take to be $2$) 
should be matched by multiplying it with a suitable power of $\alpha$. 
Once the solution is properly normalized the limit is non-zero. 
We will elaborate more on this in the future, during our proposed study of the problem at the level of actions.

\section{Summary and future works}

In this paper we solve a bilocal model up to second order in perturbation theory. The model we introduce is the simplest that fulfills the requirements of action-at-a-distance {\it and} 
short range non-local effects. A subset of the linear and second order bilocal contributions are found to encode those of gravity in the Lorentz gauge, a result that supports the idea that full gravity is encoded in bilocal theories. This claim can be extended to any non-local theory as long as it has a bilocal effective regime.\\

The comparison we have made between bilocal theories and gravity involves perturbative solutions of both theories, which is why we needed to fix the gauge in the gravity side. We believe that, in the next step, the statement that bilocal theories encode gravity should be investigated at the level of actions. Finding the way bilocal actions relate to gravity actions would be extremely clarifying. Some future lines of research regarding the bilocal action are 

\begin{itemize}

\item 
A salient feature of a gravity theory is diffeomorphism invariance. Therefore, a bilocal action must recover that symmetry in certain limit. Although we do not know details yet, it seems likely that the bilocal field needs to be promoted to a maybe multilocal) Yang-Mills field with gauge symmetry, say, $U(N)$. That symmetry could provide (maybe at large $N$) a notion of invariance under diffeomorphisms. This development would need a deeper understanding of non-local theories. For example, it is not clear  how to even gauge the bilocal theory, since the conventional method of `gauging' involves a notion of locality.  

\item 
The bilocal theory could be quantized using the path integral formalism. Bilocality is expected to render the theory divergence-free. So, an understanding of how bilocal and gravity actions relate to each other could bring some finite quantum gravity computations. 

\item 
The bilocal action should tell us the way the bilocal field couples to matter. Should the other fields be local? Maybe bilocal as well? Besides, one would want to consider the presence of a cosmological constant. Notice that the bilocal equations we have studied in this article will hold if we change $\eta_{\mu\nu}\to g^{(A)dS}_{\mu\nu}$ for index contraction. In the case of AdS, additional boundary conditions will be needed. 

\end{itemize}

There are interesting questions regarding bilocal theories themselves, regardless of their relation with gravity. For instance, does a generic non-local theory with a fundamental length scale have a bilocal effective regime? Under which circumstances? Not much is known about generic non-local theories, but if we could prove that they behave as bilocal theories in some regime, then we could study universal features of non-locality. Particularly, if the claim we make in this article is true, gravity and its uniqueness could be understood as a universal feature of {\it any} non-local theory.  \\

Other questions can point at the relation between bilocal theories and string theory. Specifically, we wonder if string theory has an effective bilocal regime reached as we integrate out the degrees of freedom of the string (turning it into a spring) while keeping its two end points. Note that the relation of string theory and multilocal theories was investigated in \cite{T4}, where the Nambu-Goto action was recovered as the multilocal theory depends on infinite points.\\

It is possible that the bilocal picture in relation with gravity brings modifications to General Relativity. Specifically those modifications that have to do with higher derivative terms in the gravity action. That is reasonably expected since the bilocal field is equivalent to a an infinite tower of high spin fields. It will be interesting to explore that connection and find possible deviations (perhaps testable) of General Relativity due to bilocality.

\vspace{0.2cm}
\noindent
{\bf Acknowledgment}

\noindent
We thank J.L. Cortes, M. Asorey, A. Segui, J. Pereira and Rafael Sorkin 
for useful discussions. We thank the anonymous Referees for useful comments which has helped 
in improving the paper. 
This work is supported by the 
Natural Sciences and Engineering Research Council of Canada and the 
University of Lethbridge. 

\appendix
\section{Gaussian integral}\label{Gauss}
Let us calculate 
\begin{equation}
I(X,r)=\int \bar{\Phi}^{(1)}(x,z)\bar{\Phi}^{(1)}(z,y)dz
\end{equation}
with 
\begin{equation}
\Phi^{(1)}(X,r)=e^{iPX}r^{\mu}r^{\nu}\bar{C}_{\mu\nu}\text{ exp}\Big[-\frac{\alpha^2}{2}\eta_{\mu\nu}r^{\mu}r^{\nu}\Big].
\end{equation}
Remember that $r$ is Euclidean, so $r^2=r^{\mu}r^{\nu}\eta_{\mu\nu}\geq 0$. 
The product of two vectors is always Minkowskian, that is, $v\cdot w=v^{\mu}\eta_{\mu\nu}w^{\nu}$, although some vectors are Euclideanized, like $r$. We will keep track of the Euclideanized vectors with the subscript E. 
Now, using the variables $X$ and $r$ related to $x$ and $y$ as in (\ref{coor}) we may write:
\begin{eqnarray}
 \bar{\Phi}^{(1)}(x,z)\bar{\Phi}^{(1)}(z,y)&=&e^{iP\cdot z}e^{iP\cdot X}\text{exp}\bigg[\frac{-\alpha^2}{2}\eta_{\mu\nu}\big((x-z)_E^{\mu}(x-z)_E^{\nu}+(y-z)_E^{\mu}(y-z)_E^{\nu}\big)\bigg] \nonumber \\
&\times&(x-z)_E^{\mu}(x-z)_E^{\nu}\bar{C}_{\mu\nu}(y-z)_E^{\sigma}(y-z)_E^{\rho}\bar{C}_{\sigma\rho}.
\end{eqnarray}
We plug this into the integral and we perform the change of variables $z\to z+X$, so we have
\begin{eqnarray}
 &&I(X,r) \nonumber \\
 &=&e^{i2P\cdot X}\int e^{iP\cdot z}\bar{C}_{\mu\nu}(z-(x-y)/2)_E^{\mu}(z-(x-y)/2)_E^{\nu}\bar{C}_{\sigma\rho}(z+(x-y(/2)_E^{\sigma}(z+(x-y)/2)_E^{\rho}\nonumber \\
&&\times \text{ exp}\bigg[\frac{-\alpha^2}{2}\eta_{\mu\nu}\big((z-(x-y)/2)_E^{\mu}(z-(x-y)/2)_E^{\nu}+(z+(x-y)/2)_E^{\mu}(z+(x-y)/2)_E^{\nu}\big)\bigg]dz\nonumber \\
&&=e^{i2P\cdot X}e^{\frac{-\alpha^2}{4}r^2}\int e^{iP\cdot z}e^{-\alpha^2 z_E^2}\bar{C}_{\mu\nu}(z-(x-y)/2)_E^{\mu}(z-(x-y)/2)_E^{\nu} \nonumber \\ &&\times \bar{C}_{\sigma\rho}(z+(x-y)/2)_E^{\sigma}(z+(x-y)/2)_E^{\rho}dz.\nonumber
\end{eqnarray}

The integral is over the whole space, so $z_{\mu}\in (-\infty, \infty)$. It is a multivariable Gaussian integral and can be solved analytically. The result is 
\begin{eqnarray}\label{gaussianintegral}
I(X,r)&=&e^{i2PX}e^{\frac{-\alpha^2}{4}r^2}\frac{\pi^2}{\alpha^8}e^{-\frac{p^2}{2\alpha^2}}\bigg[c^2\Big(\frac{p^4}{\alpha^4}-6\frac{p^2}{\alpha^2}+3\Big)+\frac{c}{2}C_{\mu\nu}r^{\mu}r^{\nu}(\alpha^2-p^2) \nonumber \\
&+&\frac{c^2}{2}(Pr)^2\big(1-\frac{\alpha^2}{p^2}\big) +\frac{c^2}{16}(Pr)^4\frac{\alpha^4}{p^4}+\frac{1}{2}\text{Tr}(C^2)-\frac{1}{2}\alpha^2C^2_{\mu\nu}r^{\mu}r^{\nu}\nonumber \\
&+&\frac{1}{16}\alpha^4(C_{\mu\nu}r^{\mu}r^{\nu})^2+\frac{c}{8}\frac{\alpha^4}{p^2}C_{\mu\nu}r^{\mu}r^{\nu}(Pr)^2\bigg].
\end{eqnarray}
The integral that appears in the model is a slight modification of $I(X,r)$. It is
\begin{equation}
 I'(X,r)=\int \bar{\Phi}^{(1)}(x',z)\bar{\Phi}^{(1)}(z,y')dz,
\end{equation}
with the variables
\begin{equation}
x'+y'=x+y,\quad x'-y'=2(x-y)\to X'=X, \quad r'=2r.
\end{equation}
It is straightforward to see that $I'(X,r)=I(X,2r)$. Having this into account, we can write the RHS of (\ref{secondmodel}) as
\begin{eqnarray}\label{integral}
&&e^{\frac{\alpha^2}{2}r^2}\int \bar{\Phi}^{(1)}(x',z) \bar{\Phi}^{(1)}(z,y') dz\nonumber \\
&&=e^{i2PX}e^{\frac{-\alpha^2}{2}r^2}\frac{\pi^2}{\alpha^8}e^{-\frac{p^2}{2\alpha^2}}\bigg[c^2\Big(\frac{p^4}{\alpha^4}-6\frac{p^2}{\alpha^2}+3\Big)+2cC_{\mu\nu}r^{\mu}r^{\nu}(\alpha^2-p^2) \nonumber \\
&&+2c^2(Pr)^2\big(1-\frac{\alpha^2}{p^2}\big) +c^2(Pr)^4\frac{\alpha^4}{p^4}+\frac{1}{2}\text{Tr}(C^2)-2\alpha^2C^2_{\mu\nu}r^{\mu}r^{\nu}\nonumber \\
&&+\alpha^4(C_{\mu\nu}r^{\mu}r^{\nu})^2+2c\frac{\alpha^4}{p^2}C_{\mu\nu}r^{\mu}r^{\nu}(Pr)^2\bigg].
\end{eqnarray}


\begin{thebibliography}{99}

\bibitem{gross}
D. Gross and P. Mende, Nucl. Phys. {\bf B303}, 407 (1988).

\bibitem{acv}
D. Amati, M. Ciafaloni, and G. Veneziano, Phys. Lett. {\bf B216}, 41 (1989).

\bibitem{maggiore} M. Maggiore, Phys. Lett. {\bf 304}, 65 (1993).


\bibitem{DFE} S.Doplicher, K. Fredenhagen, J.E.Roberts, {\it Spacetime Quantization Induced by Classical Gravity}, Phys. Lett. B {\bf 331} (1994), 39-44.

\bibitem{Ah} D. V. Ahluwalia, {\it Quantum measurement, gravitation, and locality}, Phys. Lett. {\bf B339}, 3013 (1994).
 
\bibitem{BBS} K. Becker, M. Becker, J. Schwarz, {\it String theory and M-theory. A modern Introduction,} Cambridge University Press; 1 edition (Dec 7 2006).

\bibitem{R} C. Rovelli, {\it Loop Quantum Gravity},
Living Rev. Relativity, {\bf 11}, (2008), 5.

\bibitem{M} J. Madore, {\it The fuzzy sphere,} Clam. Quantum Grav. {\bf 9} (1992) 6947.

\bibitem{LY} J. Lee, H. S. Yang, {\it Quantum gravity from noncommutative spacetime} Journal of the Korean Physical Society (2014) 65.
 

\bibitem{emergent} 
A. Sakharov, Soviet Physics Doklady {\bf 12}, 1040 (1968);
T. Jacobson, Phys. Rev. Lett. {\bf 75}, 1260 (1995);
C. Barcelo, S. Liberati and M. Visser, Class. Quantum Grav. {\bf 18}, 3595 (2001);
M. Visser, C. Barcelo and S. Liberati, Gen. Rel. Grav. {\bf 34}, 1719 (2002);
T. Padmanabhan, Mod. Phys. Lett. A {\bf 30}, 1540007 (2015);
E. Verlinde, JHEP {\bf 4}, 29 (2011).
\bibitem{DDW} P. Diaz, S. Das and M. Walton, {\it Bilocal theory and gravity I}, arXiv:1609.08631.

\bibitem{Yu} H. Yukawa, {\it Quantum Theory of Nonlocal Fields. 1. Free Fields}, Phys.Rev. {\bf 77} (1950) 219-226.~~ H. Yukawa,
{\it Quantum Theory of Nonlocal Fields. 2: Irreducible Fields and Their Interaction}, Phys.Rev. {\bf 80} (1950) 1047-1052.~~  H. Yukawa, 
{\it Structure and Mass Spectrum of Elementary Particles. 1: General Considerations}, Phys.Rev. {\bf 91} (1953) 415.~~ H. Yukawa,
{\it Structure and Mass Spectrum of Elementary Particles. 2: Oscillator Model}, Phys.Rev. {\bf 91} (1953) 416.
\bibitem{T} T. Takabayasi, {\it Relativistic Mechanics of Confined Particles as Extended Model of Hadrons -The Bilocal Case}, Supplement of the Progress of Theoretical Physics, No. {\bf 67}, 1979.~~A. Z. Capri and C.C. Chiang, {\it Extended meson fields. An alternative to quark confinement},  Nuov Cim A {\bf 36} (1976) 331.

\bibitem{bilocal} T. Shirafuji, {\it Green's Function of Bilocal Field Equations}, Prog. Theor. Phys. (1968) {\bf 39} (4): 1047-1068.~~ T. Takabayasi, Prog. Theor, Phys. Suppl. 
{\bf E65}, 339 (1965) and papers quoted therein.~~ K. Fujimura, T. Kobayashi and M. Namiki, {\it Nucleon Electromagnetic Form Factors at High Momentum Transfers in an Extended Particle Model Based on the Quark Model}, Prog. Theor. Phys. (1970) {\bf 43} (1): 73-79.~~ T. Goto and S. Naka, {\it On the Vertex Function in the Bi-Local Field}, Prog. Theor. Phys. (1974) {\bf 51} (1): 299-308.

\bibitem{Fe} R. P. Feynman, M. Kislinger and F. Ravndal,
 {\it Current Matrix Elements from a Relativistic Quark Model}, Phys. Rev. {\bf D} 3 (1971), 2706.
 
\bibitem{jevicki1} S. R. Das and A. Jevicki, {\it Large-N Collective Fields and Holography}, Phys.Rev. {\bf D68} (2003) 044011.
\bibitem{Kam} K. Kamimura, {\it Elimination of relative time in bilocal model}, Prog. Phys. {\bf 58} (1977) 1947.
\bibitem{JJY} A. Jevicji, K. Jin, Q. Ye, {\it Collective dipole model of AdS/CFT and higher spin gravity}, J. Phys. A {\bf 44} (2011) 465402.
\bibitem{Bar} I. Bars, {\it Gauge Symmetry in Phase Space Consequences for Physics and Spacetime}, Int. J. Mod. Phys. A {\bf 25}
(2010) 5235

\bibitem{jevicki} ~~ A. Jevicki and J.Yoon, {\it Bulk from Bi-locals in Thermo Field CFT}, JHEP {\bf 1602} (2016) 090.~~ A. Jevicki, K. Suzuki and J. Yoon, {\it Bi-Local Holography in the SYK Model}, JHEP {\bf 1607} (2016) 007.~~ A. Jevicki and K. Suzuki, {\it Bi-Local Holography in the SYK Model: Perturbations}, arXiv:1608.07567.~~S. R. Das, A. Jevicki, K. Suzuki, {\it Three Dimensional View of the SYK/AdS Duality}, arXiv:1704.07208. 
  
\bibitem{dMJJR} R. de Mello Koch, A. Jevicki, K. Jin, J. P. Rodrigues, {\it AdS4/CFT3 Construction from Collective Fields}, Phys.Rev. {\bf D83} (2011) 025006.

\bibitem{APR} R. Aldrovandi, J. G. Pereira, R. da Rocha and K. H. Vu, {\it Nonlinear Gravitational Waves: Their form and Effects}, Int J Theor Phys (2010) 49: 549.  arXiv:0809.2911v2.

\bibitem{RW} R. M Wald, {\it General Relativity}, The University of Chicago Press, 1984.

\bibitem{T2} T. Takabayasi, {\it Relativistic Mechanics of Two Interacting Particles and Bilocal Theory}, Prog. Theor. Phys. {\bf 54} (1975) 563.

\bibitem{BBSu} X. Beakaert, N. Boulanger, P. Sundell, {\it How higher-spin gravity surpasses the spin two barrier: no-go theorems versus yes-go examples}, Rev. Mod. Phys. {\bf 84}, 987. arXiv:1007.0435.


\bibitem{T3} T. Takabayasi, {\it Relativistic Mechanics of Bilocal and Trilocal Structures}, Prog. Theor. Phys. {\bf 57} 329 (1977).

\bibitem{T4} T. Takabayasi, {\it Theory of Relativistic String and Multilocal Model}, 
Prog. Theor. Phys. Supplement {\bf 86} (1986) 81.
\end{thebibliography}
\end{document}